\documentclass[twocolumn,aps,prl,superscriptaddress]{revtex4-2}
\usepackage{amsmath}
\usepackage{graphicx}
\usepackage{MnSymbol}
\usepackage{hyperref}
\usepackage{fancyhdr}

\begin{document}

\title{Air leakage in seals with application to syringes}

\author{N. Rodriguez}
\affiliation{BD Medical-Pharmaceutical Systems,
1 Becton Drive, Franklin Lakes, NJ 07417, USA}
\author{A. Tiwari}
\affiliation{PGI-1, FZ J\"ulich, Germany}
\affiliation{MultiscaleConsulting, Wolfshovener str. 2, 52428 J\"ulich, Germany}
\author{B.N.J. Persson}
\affiliation{PGI-1, FZ J\"ulich, Germany}
\affiliation{MultiscaleConsulting, Wolfshovener str. 2, 52428 J\"ulich, Germany}

\begin{abstract}
We study the leakage of air in syringes with Teflon coated rubber stopper and glass barrel. The leakrate
depends on the interfacial surface roughness, the viscoelastic properties of the rubber 
and on the elastoplastic properties of the Teflon coating.
The measured leakage rates are compared to the predictions of a simple 
theory for gas flow, which takes into account both the diffusive and ballistic 
air flow, and the elastoplastic multiscale contact mechanics which determines the
probability distribution of interfacial separations. 
The theory shows that the interfacial air flow (leakage) channels
are so narrow that the gas flow is mainly ballistic (the so called Knudsen limit). 
The implications for container closure integrity is discussed.
\end{abstract}

\maketitle

\thispagestyle{fancy}

{\bf 1 Introduction}

All solids have surface roughness, which has a huge influence on a large number of physical
phenomena such as adhesion, friction, contact mechanics and the 
leakage of seals\cite{Ref1,Ref2,Ref3,Ref4,Ref5,Ref6,Ref7,Ref8}.
Thus when two solids with nominally flat surfaces are squeezed into contact, unless the applied
squeezing pressure is high enough, or the elastic modulus of at least one of the solids low enough,
a non-contact region will occur at the interface. If the non-contact region percolate,
open flow channels exist, extending from one side of the nominal contact region to the other side.
This will allow fluid to flow at the interface from a high fluid pressure 
region to a low pressure region. 

We consider the leakage of air at the interface between a Teflon coated rubber stopper and a glass barrel.
We have studied this problem in Ref. \cite{Lor} but here we extend that study by using a new design of the stopper
involving much higher contact pressures, where the contact is close to the percolation threshold.

Teflon (polytetrafluorethylen) and other films (e.g., ultra-high-molecular-weight polyethylene or 
ethylene-tetrafluorethylene-copolymer) are used in laminating rubber stoppers and have elastic modulus $100-1000$
times higher than the typical rubbers stoppers ($2-6 \ {\rm MPa}$). Thus,
the average interfacial separation, resulting from the surface roughness,
is much larger than for the uncoated rubber stopper, and the contact area percolation threshold may not be achieved.
Therefore, in any laminated rubber stoppers an accurate calculation of the interfacial
separations as well as design parameters (contact pressure, geometry, etc...)
is particularly important to assure container closure integrity,
low weight losses, functional performance, and no microbial ingress during the shelf life and use of the product.
One way to study the size of the most narrow constrictions in the open (percolating) channels at the
stopper-barrel interface is by measuring the air leakage rate in syringes (with closed needle).

\vskip 0.3cm
{\bf 2 Experimental results}

The Teflon laminated rubber stopper used in this study has three ribs 1-3 which contact the glass
barrel. The ribs have (half) circular cross-section with the front and middle ribs with the 
diameter $0.4 \ {\rm mm}$, and back rib 3 with the diameter  $0.8 \ {\rm mm}$.
We first determine the width and the average contact pressure acting
in the contact region. We also study the surface topography of the
Teflon surface on the ribs. Finally we present the results of the air leakage experiments.

\begin{figure}
        \includegraphics[width=0.4\textwidth]{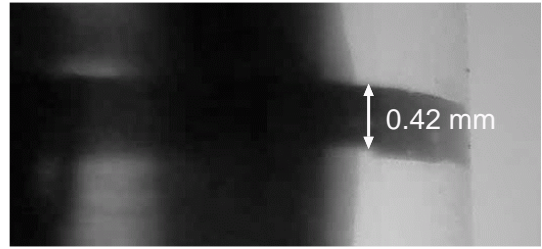}
        \caption{\label{BothInGlass.eps}
Optical picture of the first rib contact region for Teflon laminated rubber stopper in a glass
barrel. The front and the middle ribs, 1 and 2, are nominally identical and with contact width
$w \approx 0.42 \ {\rm mm}$ in the axial direction. The third outer rib 3 is wider
but the contact pressure much lower,
and this rib has only a negligible influence on the air leakage rate.
}
\end{figure}

\vskip 0.1cm
{\bf Nominal contact width $w$}

Using an optical microscope 
we have measured the width in the axial direction 
of the contact regions between the 
Teflon laminated rubber stopper and the glass barrel (see Fig. \ref{BothInGlass.eps}).
The two inner ribs, 1 and 2, are nominally identical and with the contact width
$w \approx 0.42 \ {\rm mm}$. The third (outer) rib 3 is wider
but the contact pressure is much lower, and this rib has only a small influence on the air leakage rate.
Note that the pressure is so high as to fully flatten the rib 1 and 2. We will assume below that the contact pressure
is Hertzian-like but this is only a rough estimation because of the large deformations involved.

The old design of the stopper studied in Ref. \cite{Lor} had two ribs, the inner rib with
rectangular cross section with the contact width $\approx 1.2 \ {\rm  mm}$ and back rib 
circular with the diameter $1 \ {\rm mm}$. In this case the average contact pressure ($\sim 1 \ {\rm MPa}$) is much lower
than for the new design. 

\begin{figure}
        \includegraphics[width=0.48\textwidth]{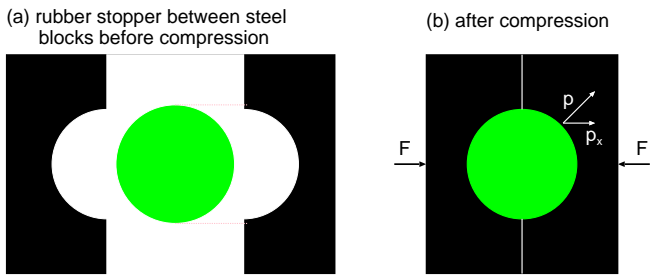}
        \caption{\label{Compress.eps}
Experimental method used to determine the radial force squeezing the stopper against the barrel.
Two steel blocks with a cylinder cavity with the same radius as the inner radius $R$ of the barrel
is squeezed together, with the rubber stopper inserted, with such a force $F$ that the gap just closes without
any repulsive force from the steel-steel contact. The steel surfaces are lubricated so negligible
shear stress occur in the contact between the stopper and the steel surface. The average 
pressure $p_0$ acting in the contact region between the stopper and the steel surface is 
determined by $p_0 D w =F$, where $w$ is the width in the axial direction of the contact region
and $D=2R$ the inner diameter of the barrel.             
}
\end{figure}

\vskip 0.1cm
{\bf Average contact pressure $p_0$}

We have measured the average contact pressure between the stopper and the barrel using the
set-up shown in Fig. \ref{Compress.eps}.
Two steel blocks with a cylinder cavity with the same radius as the inner radius $R$ of the barrel
is squeezed together with the stopper inbetween, with such a force $F$ that the gap just closes without
any repulsive force at the steel-steel contact. The steel surfaces are lubricated so negligible
shear stress occurs in the contact between the stopper and the steel surface. 
If  $w$ is the width in the axial direction of the contact region and if
$p_0$ denotes the average pressure acting in the contact region between the stopper and the steel surface 
then
$$F= \int_{-\pi/2}^{\pi/2} d\phi \ w R p_0 {\rm cos}\phi = 2wRp_0 .$$
Thus $p_0=F/wD$ where $D=2R$ is the inner diameter of the barrel.             
Using this equation we can determine the average pressure acting on each rib by removing the other ribs.
In the present case this shows that for the inner two ribs 1 and 2 (which are nominal identical with
contact width $w\approx 0.42 \ {\rm mm}$), the (average) contact pressure $p_0 \approx 3.2 \ {\rm MPa}$,
and the maximal (Hertz) contact pressure $p_{\rm max} = 3 p_0/2 \approx 4.8 \ {\rm MPa}$,
where we have used $D=6 \ {\rm mm}$. The third
(outermost) rib 3 is wider but the contact pressure much smaller, so this rib
does not affect the air leakage rate.

\begin{figure}
        \includegraphics[width=0.4\textwidth]{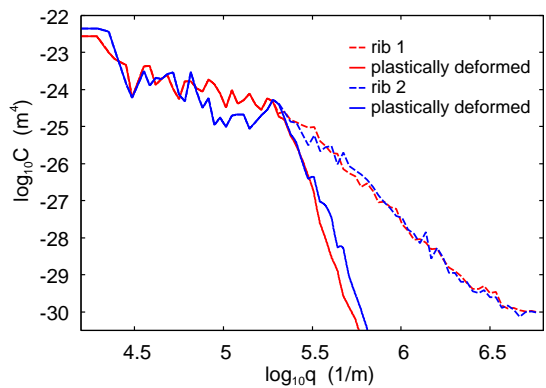}
        \caption{\label{1logq.2logC.13.14.eps}
The surface roughness power spectra obtained from stylus topography measurements on the inner rib
1 (red) and the next inner rib 2 (blue). The solid lines are the effective power spectra of the plastically deformed 
Teflon surfaces using the Teflon penetration hardness $\sigma_{\rm P} = 13 \ {\rm MPa}$.        
}
\end{figure}

\vskip 0.1cm
{\bf Surface roughness power spectrum}

The most important information about the surface topography of a rough surface is the surface roughness power spectrum.
For a one-dimensional (1D) line scan $z=h(x)$ the power spectrum is given by
$$C_{\rm 1D} (q) = {1\over 2 \pi} \int dx \ \langle h(x) h(0) \rangle e^{i q x} \eqno(1)$$
where $\langle .. \rangle$ stands for ensemble averaging.
For surfaces with isotropic roughness the 2D power spectrum $C(q)$ can be obtained directly 
from $C_{\rm 1D} (q)$ as described elsewhere \cite{Nayak,Carbone}.
For randomly rough surfaces, all the (ensemble averaged) information about the surface 
is contained in the power spectrum $C(q)$ (see Ref. \cite{Ref5,Ref6}). For this reason
the only information about the surface roughness which enter in contact mechanics
theories (with or without adhesion) is the function $C(q)$.
Thus, the  (ensemble averaged) area of real contact, the interfacial stress distribution, and the
distribution of interfacial separations, are all determined by $C(q)$.

We have measured the surface roughness profile using a stylus instrument 
[Mitutoyo Portable Surface Roughness Measurement Surftest SJ-410 with a
diamond tip with the radius of curvature $r_0 =1 \ {\rm \mu m}$, and with the tip-substrate repulsive force $F_{\rm N} = 0.75 \ {\rm mN}$
and the tip speed $v=50 \ {\rm \mu m/s}$],
and calculated the surface roughness
power spectrum as described in detail elsewhere\cite{Ref6}. The dashed lines in Fig. \ref{1logq.2logC.13.14.eps}
shows the surface roughness power spectra obtained from stylus topography measurements on the rib 1 
(red) and rib 2 (blue). The solid lines are the effective power spectra of the plastically deformed 
Teflon surfaces obtained as described in Ref. \cite{Heat} and summarized below. In the calculation we have used 
the rubber modulus $E=4.6 \ {\rm MPa}$ and Poisson ratio $\nu = 0.5$, the Teflon film thickness $d=15 \ {\rm \mu m}$,
the Teflon modulus $E=500 \ {\rm MPa}$ and Poisson ratio $\nu = 0.4$, and the 
Teflon penetration hardness $\sigma_{\rm P} = 13 \ {\rm MPa}$. The penetration hardness of Teflon (without filler) is typically in the
range $20-30 \ {\rm MPa}$, but we interpret the lower penetration hardness we use as an effective hardness as the
Teflon surface is exposed also to shear stresses and wear processes as it moves in the glass barrel, which smooth
the surface. In addition, some plastic flow occur already when the contact pressure is 
well below the penetration hardness\cite{Jonson}. In fact, it has been observed that Teflon start 
to flow plastically around $13 \ {\rm MPa}$ in wear experiments\cite{Florida}.

Here we note that the Persson contact mechanics theory assumes randomly 
rough surface roughness. However, plastic deformation in general
result in non-random roughness. The procedure we use to obtain the power spectrum for
plastically deformed surfaces has been described in detail elsewhere\cite{Heat} but is briefly summarized here.

In elastic contact mechanics the contact area decreases continuously as we increase the magnification and observe shorter
wavelength roughness. Hence, when a solid with surface roughness is squeezed against a flat rigid surface with the force $F_{\rm N}$
the solid may deform elastically in the contact regions observed at low magnification,
where the contact area $A$ is large and the contact stress $F_{\rm N}/A$ low, but
plastically above some critical magnification which depends on the penetration hardness.
We take this plastic deformation into account by smoothing (or removing) the short wavelength roughness.
We do this in such a way that the elastic contact area of the plastically deformed surface will be the same
(as a function of magnification) as that obtained using the Persson elastoplastic contact mechanics theory\cite{Persson2,PlastPRL}.
But in order for the surface to be randomly rough one must smooth the surface
everywhere, i.e., also in the non-contact area. We have argued
before\cite{Ref5,skew} that this has only a small influence of the interfacial separation in the open flow channels
relevant for the fluid leakage problem. Nevertheless, the power spectrum obtained this way is an effective
power spectrum designed for special purpose, and it cannot be compared to the real power spectrum obtained
from the surface topography of the plastically deformed surface, as discussed in our earlier studies\cite{PlastAl}.

\begin{figure}
\includegraphics[width=0.45\textwidth]{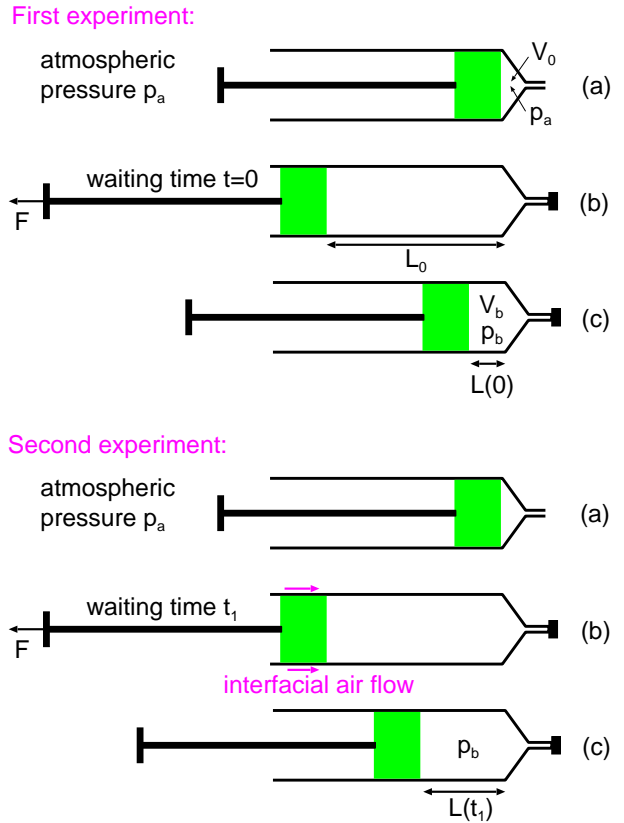}
\caption{\label{two3sy.eps}
Experiments performed in order to measure the air leak-rate of syringes (see text for details).
}
\end{figure}

\vskip 0.1cm
{\bf Leakage rate}

We have measured the air leakage in 15 syringes using the same procedure as in Ref. \cite{Lor}
and summarized in Fig. \ref{two3sy.eps}. We first assemble
the barrel-stopper in empty configuration with the stopper pushed to the end of the barrel, 
resulting in a small volume $V_0$ of gas in the syringe at atmospheric pressure.
Next the needle is closed so no air can penetrate into the syringe from the needle side, and the stopper is
pulled back (retracted) to full fill position resulting in a volume $L_0A_0$ of gas at 
low pressure. In the first experiment the pull-force is immediately removed, which
results in the stopper moving to a new position $L(0)$ where the 
gas pressure $p_{\rm b}$ is such that the pressure force $(p_{\rm a}-p_{\rm b})A_0$ (due to the difference in the
gas pressure outside and inside the barrel) is equal to the stopper-barrel friction force. Next we repeat the experiment
except now the stopper is kept in the pulled back (retracted) position for some time $t_1$.
This results (due to air leakage at the barrel-stopper interface) in an increase in the air pressure, and when the
pull force is removed after some time $t_1$ the stopper will move to a new position with $L(t_1)>L(0)$.
The volume increase $\Delta V = [L(t_1)-L(0)]A_0$ is due to the air leakage into the syringe.
However, the air pressure in the volume $\Delta V$ is not the atmospheric air pressure $p_{\rm a}$ but is equal
to $p_{\rm b}$. Thus, the $\Delta V$ correspond to a volume 
$\Delta V_{\rm a} =\Delta V p_{\rm b}/p_{\rm a}$ of air of atmospheric pressure. 
Since no leakage is assumed to occur during the first experiment we have $p_{\rm a} V_0 =p_{\rm b} V_{\rm b}$
so that $p_{\rm b} /p_{\rm a} = V_0/V_{\rm b}$. Hence the volume of air of atmospheric pressure
leaking into the syringe per unit time equal
$$\dot V = {\Delta V_{\rm a}\over t_1} = 
{\Delta V \over t_1} {p_{\rm b} \over p_{\rm a}} = {\Delta V \over t_1}  {V_0 \over V_{\rm b}}$$

It is easy to study the leakage for each rib separately by introducing a thin 
cut in the other ribs through which the air (or fluid) can move with negligible
resistance.

In our experiments, the pressure difference between inside and outside the syringe is about $1 \ {\rm bar}$,
and the average air leakage rate about $2.2 \times 10^{-4} \ {\rm mm^3/s}$.
We repeated the test in five of the fifteen syringes obtaining very similar leakage result as in the initial measurements.
In an earlier study (see Ref. \cite{Lor}) with a different design of the Teflon coated rubber stopper, where the
average contact pressure was much smaller, we observed larger air leakage rates, of order
$4.8 \times 10^{-3} \ {\rm mm^3/s}$, i.e. about a factor of 20 higher than for the new design.

\begin{figure}
        \includegraphics[width=0.45\textwidth]{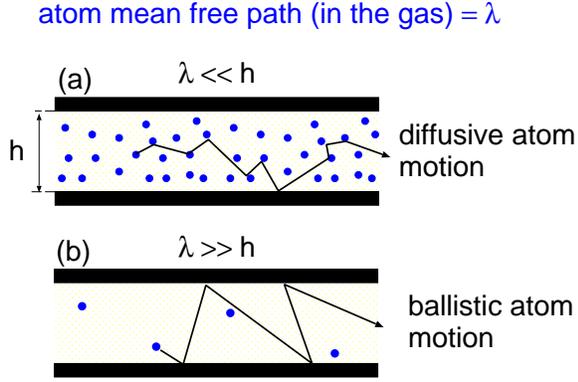}
        \caption{\label{Ballistic.eps}
Diffusive (a) and ballistic (b) motion of the gas atoms (e.g. He) in the critical junction.
In case (a) the gas mean free path $\lambda$ is much smaller than the gap width $h=u_{\rm c}$ and the gas
molecules makes many collisions with other gas molecules before a collision with the solid walls.
In the opposite limit, when $\lambda >> u_{\rm c}$ the gas molecules makes many collisions with the solid wall before colliding
with another gas molecule. In the first case (a) the gas can be treated as a
(compressible) fluid, while a kinetic approach is needed in case (b).}
\end{figure}

\begin{figure}
        \includegraphics[width=0.45\textwidth]{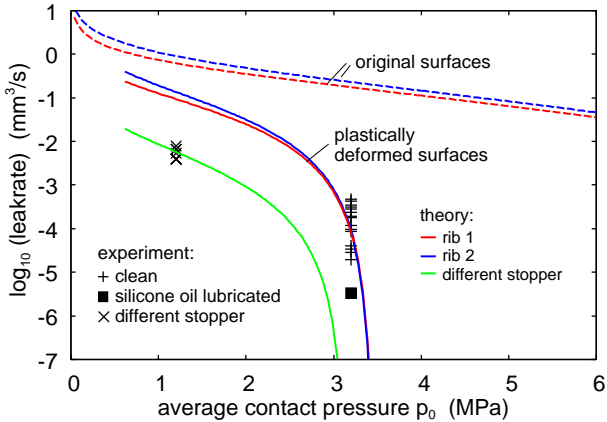}
        \caption{\label{1pressure.2logLeakage.eps}
Solid and dashed lines are the calculated leakage rate as a function of the average contact pressure
using the plastically deformed surface (solid lines) and the original surface (dashed lines). The + symbols are the
measured leakage rates using different (but nominally identical) clean syringes. The square symbol is for 
the same type of syringe but with the glass barrel lubricated by silicone oil. 
The $\times$ symbols are for a different Teflon coated rubber
stopper where the rib contact regions are much wider and the contact pressure much smaller (see Ref. \cite{Lor}).}
\end{figure}

\begin{figure}
        \includegraphics[width=0.4\textwidth]{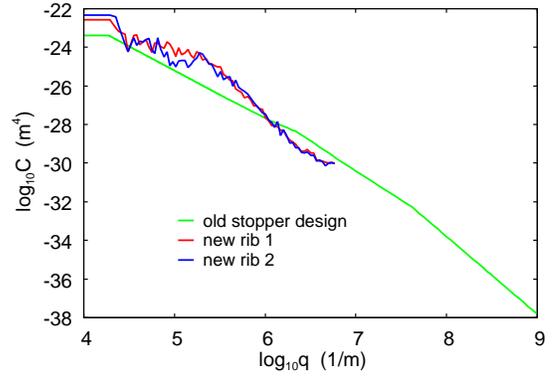}
        \caption{\label{1loq.2logC.new.old.eps}
The surface roughness power spectra obtained from stylus topography measurements on the inner rib 1
(red) and the next inner rib 2 (blue). The green line is the power spectrum for another design of the laminated rubber
stopper with much wider rib contact regions. 
For this stopper both engineering stylus, optical and Atomic Force Microscopy
topography was perform and the green line is a fit to all the measured data (see Ref. \cite{Lor}).
}
\end{figure}

\vskip 0.3cm
{\bf 3 Analysis of experiments}

The leakrate of fluids at interfaces between solids with random roughness can be calculated
using the critical junction theory or the Bruggeman effective medium theory as described in detail
elsewhere\cite{Boris,LP1,Dapp,Dapp1, More1,More2,More3,Scarag1,Scarag2}. 
In the critical junction theory it is assumed that all the pressure drop in the fluid
occurs at the most narrow constrictions along the biggest open (percolating) flow channels.
The more accurate effective medium theory includes all flow channels in an approximate way
but both theories gives usually very similar results. The probability distribution
of interfacial separations, which enter in the theory for the leakage rate, is determined using
the Persson contact mechanics theory\cite{Persson2,aaa,Carbone1,LP1,Prodanov}. 
In the present case, with the rubber stopper covered by a thin
Teflon film, one must include the plastic deformations of the Teflon surface roughness profile,
see Sec. 2 and Ref. \cite{Heat}.

For the leakage of fluids one can usually assume laminar flow of Newtonian fluid characterized by
a viscosity $\eta$. This description is also valid for gases if the surface separation at the critical constriction is much larger than the
gas molecule mean free path $\lambda $. However, this is not the case in the present application where the
flow through the critical constriction is ballistic rather than diffusive, 
see Fig. \ref{Ballistic.eps}. In Ref. \cite{Suction,Preparation} we have presented an interpolation formula
for gas flow through a narrow pore which correctly describes both the diffusive (large pore diameter) and
ballistic (narrow pore diameter) limits. In the equation enters the viscosity $\eta$, the mean free path $\lambda $ and the average
velocity $\bar v$ of a gas molecules. 
We have used $\eta = 1.76\times 10^{-5} \ {\rm Pa s}$, $\lambda = 59 \ {\rm nm}$ and $\bar v = 470 \ {\rm m/s}$.

The solid and dashed lines in Fig. \ref{1pressure.2logLeakage.eps}
shows the calculated leakage rate as a function of the average contact pressure
using the plastically deformed surface (solid lines) and the original surface (dashed lines). The + symbols are the
measured leakage rates using different (but nominally identical) clean syringes. The square symbol is for 
the same type of syringe, but with the glass barrel lubricated by silicone oil. 
The silicone oil block air flow channels and reduces the air leakage rate, and one may have expected an even larger
reduction than observed. Note the crucial influence of the
plastic deformation which for the relevant average contact pressure $p_0 \approx 3.2 \ {\rm MPa}$ reduces the leakage rate
by a factor of $\sim 1000$. 

The maximum contact pressure for the rib contacts 1 and 2 is very close to the pressure where the contact area percolate.
This result in the large fluctuations (by a factor of nearly 100) 
in the measured leakage rate between nominally identical syringes, and also in a
large sensitivity in the calculated leakage rate to small variations 
in the system parameters, e.g., the penetration hardness. 

The $\times$ symbols in Fig. \ref{1pressure.2logLeakage.eps} are for a different Teflon coated rubber
stopper studied in Ref. \cite{Lor},
where the rib contact regions are much wider and the contact pressure much smaller. For this stopper the Teflon coating was
slightly smoother (see Fig. \ref{1loq.2logC.new.old.eps}) than  in this study, and when this is 
taken into account the theory prediction agrees very well with the measured data. This is shown by the
green line in Fig. \ref{1pressure.2logLeakage.eps}, which was calculated using the power spectrum of the plastically deformed
surface obtained in the same way as for the rib 1 and 2. Note that the fluctuations in the measured data is much smaller than
for the new stopper design, which is consistent with the fact that the contact pressure is well below the pressure where the
contact area percolate. 

\begin{figure}
        \includegraphics[width=0.4\textwidth]{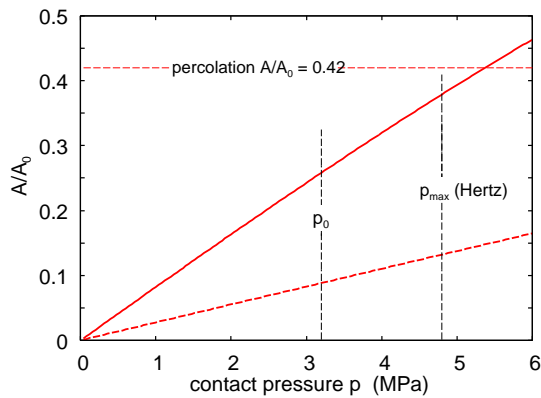}
        \caption{\label{1pressure.2Area.13.eps}
The contact area as a function of the contact pressure for rib 1. The solid line 
is for the plastically deformed surface and the dashed line for the originally (not deformed)
Teflon surface. The contact area for a randomly rough surface percolate when $A/A_0 \approx 0.42$.
The dashed lines indicate the average and the maximum contact pressure assuming Hertz-like pressure distribution.
}
\end{figure}

For a randomly rough surface the contact area percolate when $A/A_0 \approx 0.42$ (see Ref. \cite{Dapp}).
When the contact area percolate no open flow channel occurs at the interface and the leakage rate vanish.
Fig. \ref{1pressure.2Area.13.eps} shows the contact area as a function of the contact pressure for rib 1. The solid line 
is for the plastically deformed surface and the dashed line for the originally (not deformed)
Teflon surface. The dashed lines indicate the average and the maximum contact pressure assuming Hertz-like pressure distribution.
Note that for the maximum contact pressure the contact area nearly percolate.

Container closure integrity is very important 
for syringes, so that no microorganism (bacteria and viruses) can penetrate from the outside to
inside the syringe. The diameter of virus is in the range 
$\approx 20-400 \ {\rm nm}$ and it is clear that 
complete container closure integrity would imply
that the most narrow junction (denoted critical junction) in the largest open (non-contact)
interfacial channel should be at most $20 \ {\rm nm}$. When this condition is satisfied, the
fluid leakage is also negligible.   

\begin{figure}
        \includegraphics[width=0.45\textwidth]{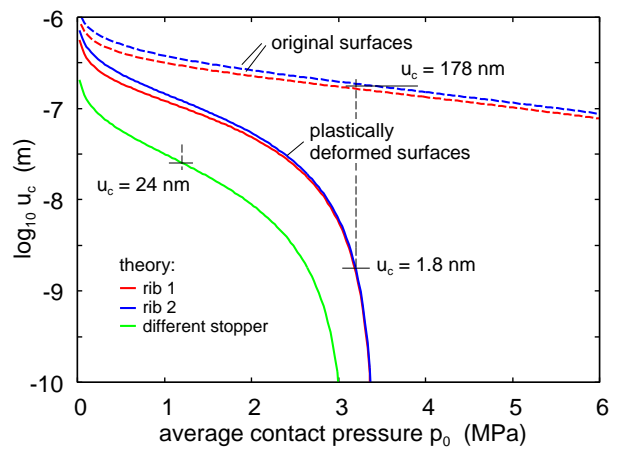}
        \caption{\label{1pressure.2loguc.all.eps}
The calculated surface separation at the critical constriction as a function of the average contact pressure
using the plastically deformed surface (solid lines) and the original surface (dashed lines). 
The critical constriction is the most narrow constriction along the biggest open 
flow channels at the Teflon-glass interface.}
\end{figure}

Fig. \ref{1pressure.2loguc.all.eps}
shows the calculated surface separation at the critical constriction as a function of the average contact pressure
using the plastically deformed surface (solid lines) and the original surface (dashed lines). 
The critical constriction is the most narrow constriction along the biggest open 
flow channels at the Teflon-glass interface. The separation of the surfaces at the
critical constriction is $\sim 2 \ {\rm nm}$, which implies container closure integrity.

The theory predict that the separation between the surfaces at the most narrow constrictions along the 
biggest fluid flow channels is so small ($\sim 24 \ {\rm nm}$) that also for 
the old design no bacterial ingress is possible, which has been confirmed experimentally.

\vskip 0.3cm
{\bf 4 Discussion}

The study presented in this paper and in Ref. \cite{Lor,Al} shows the importance of plastic flow
in some applications to seals. The good agreement found here, and in Ref. \cite{Lor},
support the procedure we use to include the plastic
deformation. The role of plastic deformation was studied for metallic seals in Ref. \cite{Al} where
the theory showed that the plastic flow reduce the leakage rate with a factor of $\sim 8$, resulting in water
leakage rates in good agreement with experiments.
We have shown in Ref. \cite{Pol,PlastAl} that plastic flow may 
modify the surface topography in different ways for metals and polymers, but at least for Teflon
and steel the way we include plastic flow gives good results for both systems.

We note that the interfacial separations predicted by the theory for the new and old stopper design
($2-25 \ {\rm  nm}$) are so small that gas leakage through molecular diffusion in
the elastomer material\cite{Diff,Huon} may contribute in an important way to the
measured leakage rate. In fact, for the syringe lubricated by silicone oil, 
where $\dot V \approx 2.5\times 10^{-6} \ {\rm mm^3/s}$, this might be the dominant leakage mechanism.
As a consequence the traditional way of container closure integrity 
studies (dye ingress or gas transport through the interface) is a subject of debate nowadays.

\vskip 0.3cm
{\bf 5 Summary and conclusion}

We have performed air leakage experiments for the contact between a rubber stopper, laminated with a thin Teflon film, and
a smooth cylindrical glass barrel. 
The measured leakrates where compared with theory predictions based on a theory
involving gas flow through narrow constrictions taking into account the interfacial separation and 
mean-free path of the gas molecules. 
We used the Perssons contact mechanics theory to 
calculate the probability distribution of surface separation at the stopper/glass interface,
and the Bruggeman effective medium theory to calculate the fluid 
flow resistance in the complex set of interconnected flow channels at the interface.
We found that the plastic deformation of the Teflon surface reduces the 
interfacial separation by a factor of $\sim 100$, and result in a reduction of the 
leakrate by a factor of $\sim 1000$.
The measured leakage rates are in good agreement with the theory predictions,
but exhibit large fluctuations because the contact is close to the percolation
threshold where small variations in the system parameters can generate large changes 
in the leakage rate.

\vskip 0.5cm
{\bf Acknowledgments:}

We thank Martin M\"user for useful comments on the manuscript.
We thank Lucile Gontard for checking some roughness parameters.

\end{document}